\documentclass[apj]{emulateapj-rtx4}
\usepackage{natbib}
\usepackage{multirow}



\newcommand{\pasa}{PASA}

\begin{document}

\bibliographystyle{apj}

\title{An ultraluminous Lyman Alpha emitter with a blue wing at z=6.6
\altaffilmark{1,2}}

\author{
E.~M.~Hu \altaffilmark{3},
L.~L.~Cowie\altaffilmark{3},
A.~Songaila\altaffilmark{3}, 
A.~J.~Barger\altaffilmark{4,5,3}, 
B.~Rosenwasser\altaffilmark{4},
I.~Wold\altaffilmark{6}
}

\altaffiltext{1} {Based on data collected at the Subaru Telescope, which is
operated by the National Astronomical Observatory of Japan.}
\altaffiltext{2}{The W.~M.~Keck Observatory is operated as a scientific
partnership among  the California Institute of Technology, the University
of California, and NASA, and was made possible by the generous financial
support of the W.~M.~Keck Foundation.}
\altaffiltext{3}{Institute for Astronomy, University of Hawaii,
2680 Woodlawn Drive, Honolulu, HI 96822.}
\altaffiltext{4}{Department of Astronomy, University of Wisconsin-Madison,
475 N. Charter Street, Madison, WI 53706.}
\altaffiltext{5}{Department of Physics and Astronomy, University of Hawaii,
2505 Correa Road, Honolulu, HI 96822.}
\altaffiltext{6}{Department of Astronomy, the University of Texas at Austin,
2515 Speedway Blvd., Stop C1400, Austin, TX 78712}

\slugcomment{Accepted for publication in Astrophysical Journal Letters}


\begin{abstract}

We report the detection of the most luminous high-redshift Lyman Alpha
Emitting galaxy (LAE) yet seen, with $\log L({\rm Ly}\alpha) = 43.9~{\rm ergs\
s^{-1}}$.  The galaxy -- COSMOS Ly$\alpha$1, or COLA1 -- was detected in a
search for ultraluminous LAEs with Hyper Suprime-Cam on the Subaru telescope.
It was confirmed to lie at $z=6.593$ based on a Ly$\alpha$ line detection
obtained from followup spectroscopy with the DEIMOS spectrograph on Keck2.
COLA1 is the first very high-redshift LAE to show a multi-component
Ly$\alpha$\ line profile with a blue wing, which suggests that it could lie in
a highly ionized region of the intergalactic medium and could have significant
infall.  If this interpretation is correct, then ultraluminous LAEs like
COLA1 offer a unique opportunity to determine the properties of the HII
regions around these galaxies which will help in understanding the ionization
of the $z\sim7$ intergalactic medium.

\end{abstract}



\section{Introduction}
\label{secintro}

The epoch of reionization is a key time in the evolution of the Universe, when
the intergalactic medium (IGM) transitioned from being neutral to being very
highly ionized. Determining when reionization occurred and finding the sources
of the photons responsible is one of the major goals in current observational
cosmology.
Ly$\alpha$ emission is one of the few diagnostics of this epoch.  Since it can
be modified by the radiative damping wings of a neutral IGM, both the observed
strength and the shape of the line are sensitive probes of the fraction of
neutral hydrogen in the IGM \citep[e.g.][]{haiman99,robertson10}.
There is a developing consensus that the fraction of star-forming galaxies
with Ly$\alpha$ emission (LAEs) drops rapidly from $z=6$ to $z=8$
\citep[e.g.][]{stark15,treu13,tilvi14,schenker14}, which could reflect a rapid
increase in the  neutral hydrogen fraction in the IGM with increasing
redshift.

However, theoretical modeling of the evolution of the LAE population is
extremely difficult \citep[e.g.][]{choudhury15}, and it is possible that this
evolution may be driven by changes in the LAE properties of the galaxies as
well as by IGM evolution \citep[e.g.][]{stark15}.  The galaxy itself produces
a complex  Ly$\alpha$ spectrum.  At $z\sim2-3$, Ly$\alpha$ lines generally
have red offsets but in some $30\%$ of cases show multi-component structures
with blue wings \citep{kulas12}.
The profiles can be crudely understood in terms of the velocity and structure
of infalling and outflowing gas in the galaxy \citep[e.g.][]{verhamme12}.  At
higher redshifts the IGM begins to modify the profiles
\citep[e.g.][]{dijkstra14}.  Beyond about $z\sim4$ the Ly$\alpha$ forest
begins to become optically thick \citep{songaila04} and  will scatter the blue
side of the Ly$\alpha$ line (and potentially some of the red side if there is
infall of the IGM to the galaxy) out of the line of sight.  This produces  the
asymmetric red profile that is characteristic of many of the LAEs. No
multi-component LAE profiles are seen in the $z=3-7$ LAE study of \citet{u15}
or in the $z=5.7$ and $z=6.6$ LAE samples of \citet{hu10} though there is wide
variation in the degree of skewness in the lines.  Multi-component structure
can only  be seen in  $z>>4$ galaxies if they highly ionize their surroundings
and this is most likely for the most luminous galaxies.  As the IGM becomes
substantially neutral near reionization both the red and blue sides of the
line will be suppressed by the radiative damping wings produced by the neutral
part of the IGM lying outside the galaxy's HII region. The size of this effect
depends on the neutral density in the surrounding IGM and also on the size of
the HII region \citep{haiman02}. In this case seeing the line at all depends
on the formation of a large HII region and again this is more probable for the
most luminous galaxies.

In the present paper we report the detection of the most luminous
high-redshift LAE yet seen.  This galaxy, which we call COSMOS Ly$\alpha$ 1
(COLA1), is the first high-redshift LAE ($z=6.593$) to show a multi-component
velocity structure with a  blue wing, which suggests that COLA1 may indeed lie
in a very highly ionized region of the IGM.

\newpage

\section{Data}
\label{secdata}

We have been conducting a search for ultraluminous LAEs/faint AGN at $z=6.6$
in six 4~deg$^2$ regions surrounding the COSMOS (158.082, 2.373), SSA22
(334.349, 0.2631), Lockman (160.033, 58.374), A370 (39.994, -1.586), CDF--S
(53.123, -27.800), and CDF--N (189.22, 62.234) fields (central J2000 RA, Dec
in decimal degrees in parentheses).
In each case we have obtained deep $g', i', z',$ and $ y'$ images with Hyper
Suprime-Cam \citep[HSC;][]{miyazaki12} and narrow-band 9210~\AA\ mosaicked
images with Suprime-Cam, both on the Subaru telescope. The full data set and
its implications for the behaviour of the bright end of the Ly$\alpha$
luminosity function will be described in Hu et al (2016; in preparation).
Here we describe the data in the COSMOS field, the first field for which we
have obtained complete spectroscopic follow-up with the Keck2/DEIMOS
spectrograph.

%
\begin{figure*}[ht]
\centerline{\includegraphics[width=10cm,angle=90]{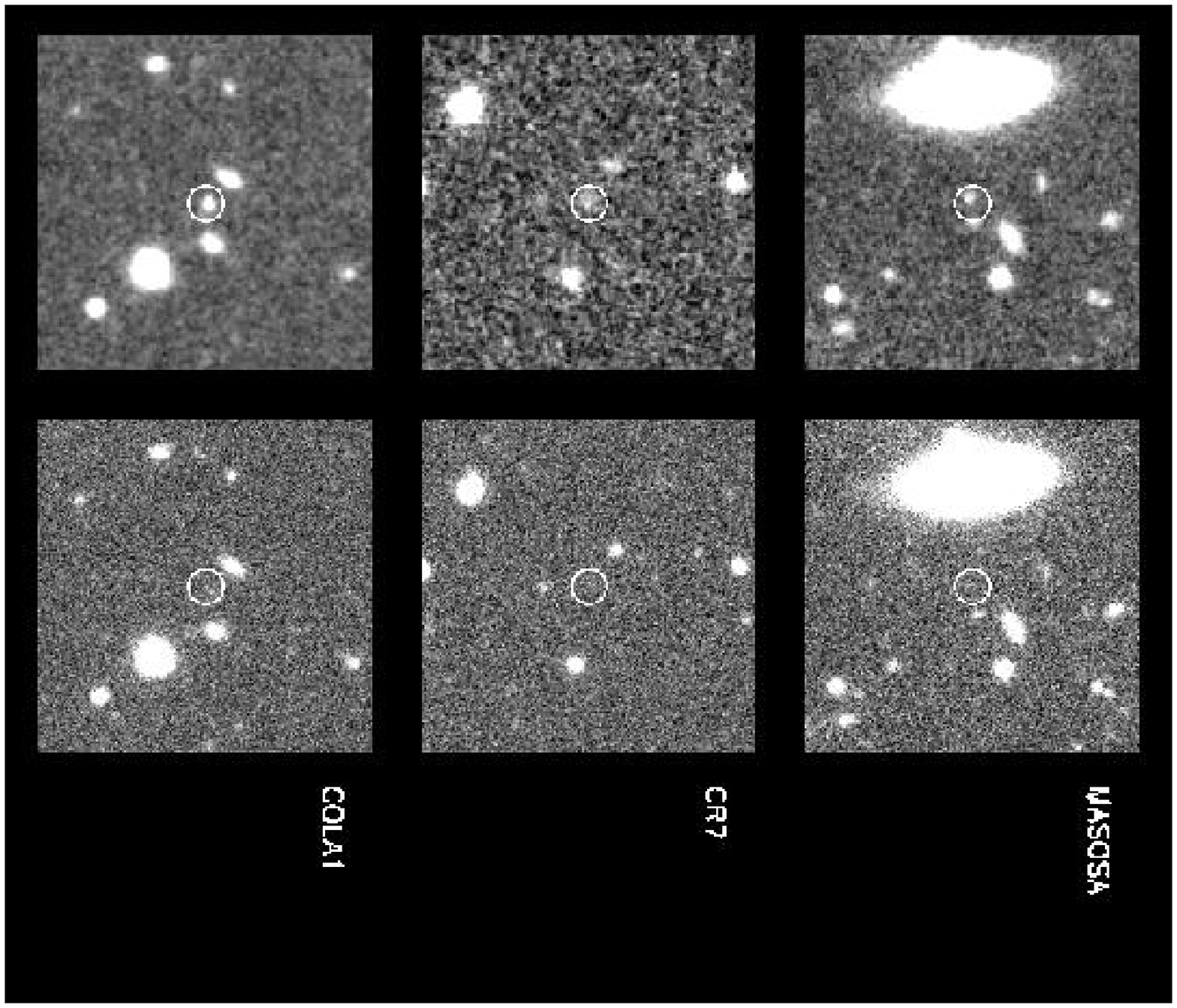}
\includegraphics[width=9cm,angle=0]{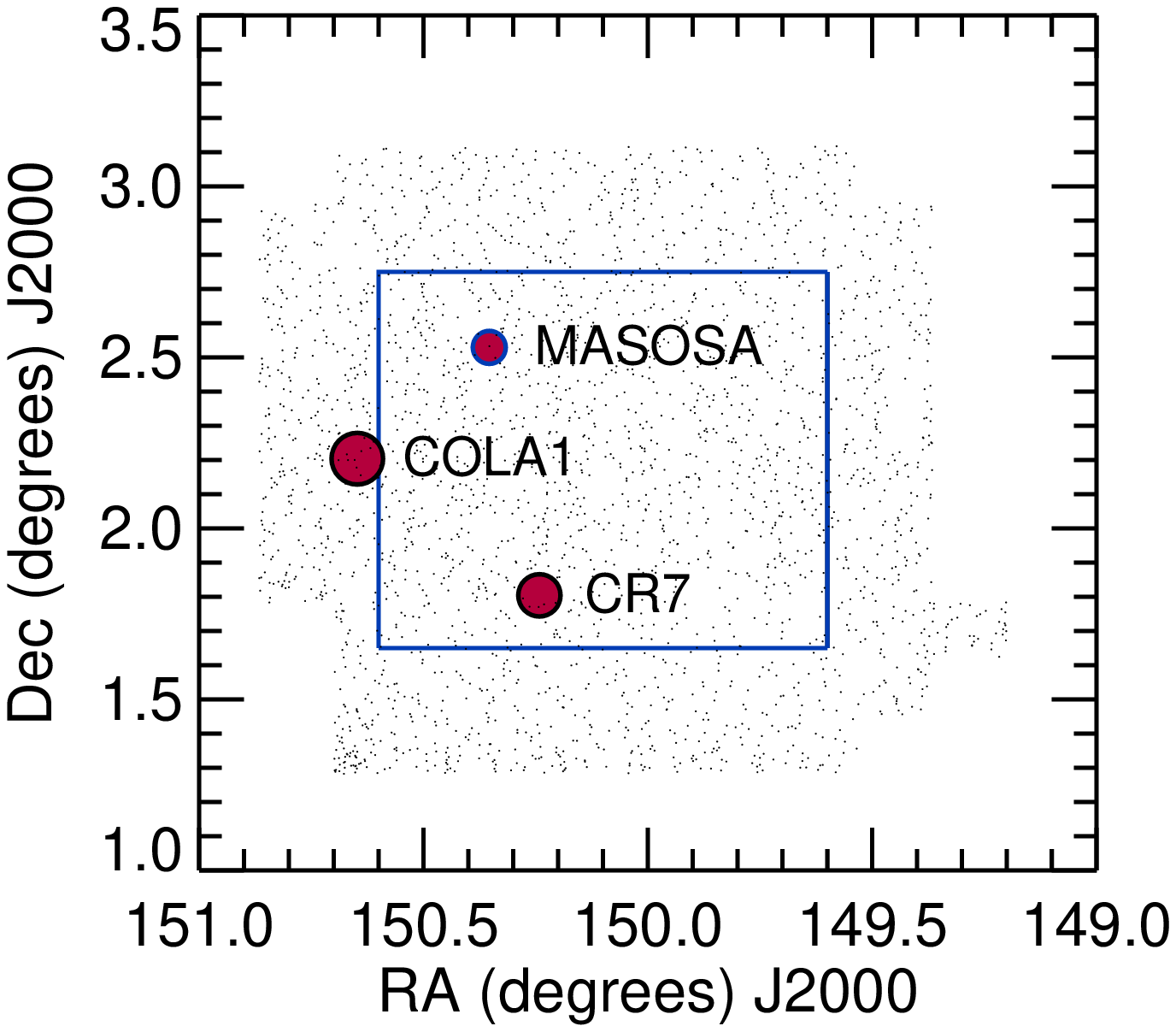}}
\caption{{\it Left}\/: Images (left -- NB921, right --  $z^{\prime}$) of the
three brightest LAEs in our extended COSMOS field.  The LAE is circled in each
image. Thumbnails are $30^{\prime\prime}$ on a side. All three galaxies are
physically extended.  COLA1 has an angular size of  $\sim 0.2^{\prime\prime}$.
{\it Right}\/: Location of these objects in the region around the COSMOS field
(red circles). The black shading shows the region covered by the present
narrow-band observations while the blue square shows the approximate region
covered by \citet{sobral15}.  COLA1 lies outside this latter region.
\label{show_images}
}
\end{figure*}

The HSC observations of the COSMOS field were made with four
overlapped HSC fields providing full coverage of  an area with  
$\sim2.1^{\circ}$  on a side. The data was obtained in March 2014
and January 2015. For each individual field we made two 5-point dither
observations with an $11^\prime$  throw. This was repeated in 4
orthogonal rotations (0,90,180,270) allowing removal of
stellar diffraction spikes. We obtained observations in
$g'$ (31 min/pixel) $i'$ (90 min,  variable transmission) 
$z'$ (30 min) and
$y'$ (85 min, variable transmission). 
Exposure times in the overlapped regions are larger.
The seeing ranged from $0.75^{\prime\prime}$ in most of
the $z'$ band to $1.1-1.2^{\prime\prime}$  in the $y'$. 
The data was reduced with
the HSC pipeline (http://hsca.ipmu.jp/public/pipeline/installation.html).
The corresponding narrow-band observations
were taken with the NB921 filter on the Suprime camera
(center, 9183\AA; FWHM, 132\AA.) 144
fields were observed with a $10^\prime$  step over 2 nights in Dec
2014. The total exposure per pixel was 900s. The
observations were reduced with  our standard pipeline  
\citep{capak04}, 
then
combined into a single image, astrometrically
matched to the HSC images. All narrow-band
data was photometric with seeing $\sim 0.9^{\prime\prime}$.

We next formed a catalog of objects over the uniformly
covered portions of the NB921 image to a limiting magnitude
of NB921(AB)$=24.25$ ($3^{\prime\prime}$ diameter aperture magnitude)
corresponding to S/N $>5~\sigma$  throughout the
area. The final narrow-band 
area coverage of 3 deg$^2$ is shown
in the right panel of Figure~1, compared with the
smaller narrow-band field of 
\citet{sobral15}.
We next searched for $z=6.6$ LAE candidates with ($z'$-NB921)$>1$ 
that were not detected above a 2$\sigma$ level
 in  $g'$ and $i'$. The narrow-band
and broad-band images of each of the candidates were next visually
inspected and artifacts (moving objects, ghosts, and contamination
by nearby bright objects) removed. All of the objects
also lie within the Ultravista field and we also visually inspected
the images of 
\citet{mccracken12}.
We were left with a set
of three objects, shown in Figure~1.
Two of these, MASOSA, and CR7, were previously known 
and spectroscopically confirmed by 
\citet{sobral15}
and 
\citet{matthee15}.
However COLA1, the most luminous
of the three,  lies outside the field observed by
these authors (Figure~1). We compare the photometric properties
of COLA1 and CR7 in Table~1 based on the present data.
Both CR7 and COLA1 lie at the long-wavelength edge of
the NB921 filter and we must allow for the falling transmission
in the filter when  converting the narrow-band magnitude to
a flux. Allowing for the relative transmission factors, the
imaging data shows COLA1 to be about $20\%$ more luminous than CR7.
%
\begin{figure*}[ht]
\hskip 0.6cm
\includegraphics[width=16cm,angle=0]{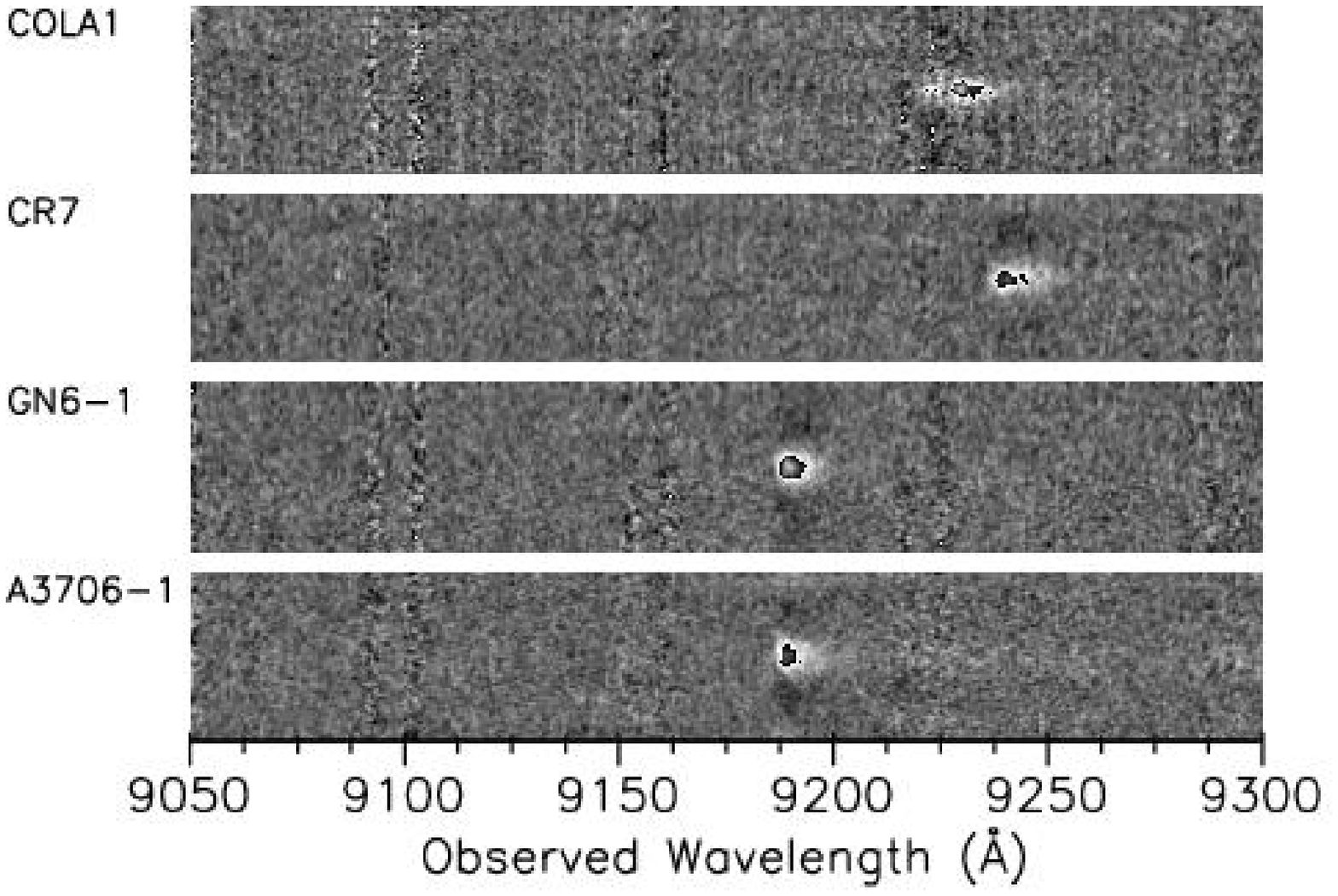}
\caption{Two-dimensional spectral images of COLA1 (top) and
CR7 (middle) compared with the brightest $z=6.6$ LAE in the
GOODS-N (GN6-1) and A370 (A3706-1) fields of \citet{hu10}
(bottom).
\label{2d-images}
}
\end{figure*}

\begin{figure}
\includegraphics[width=10cm,angle=0]{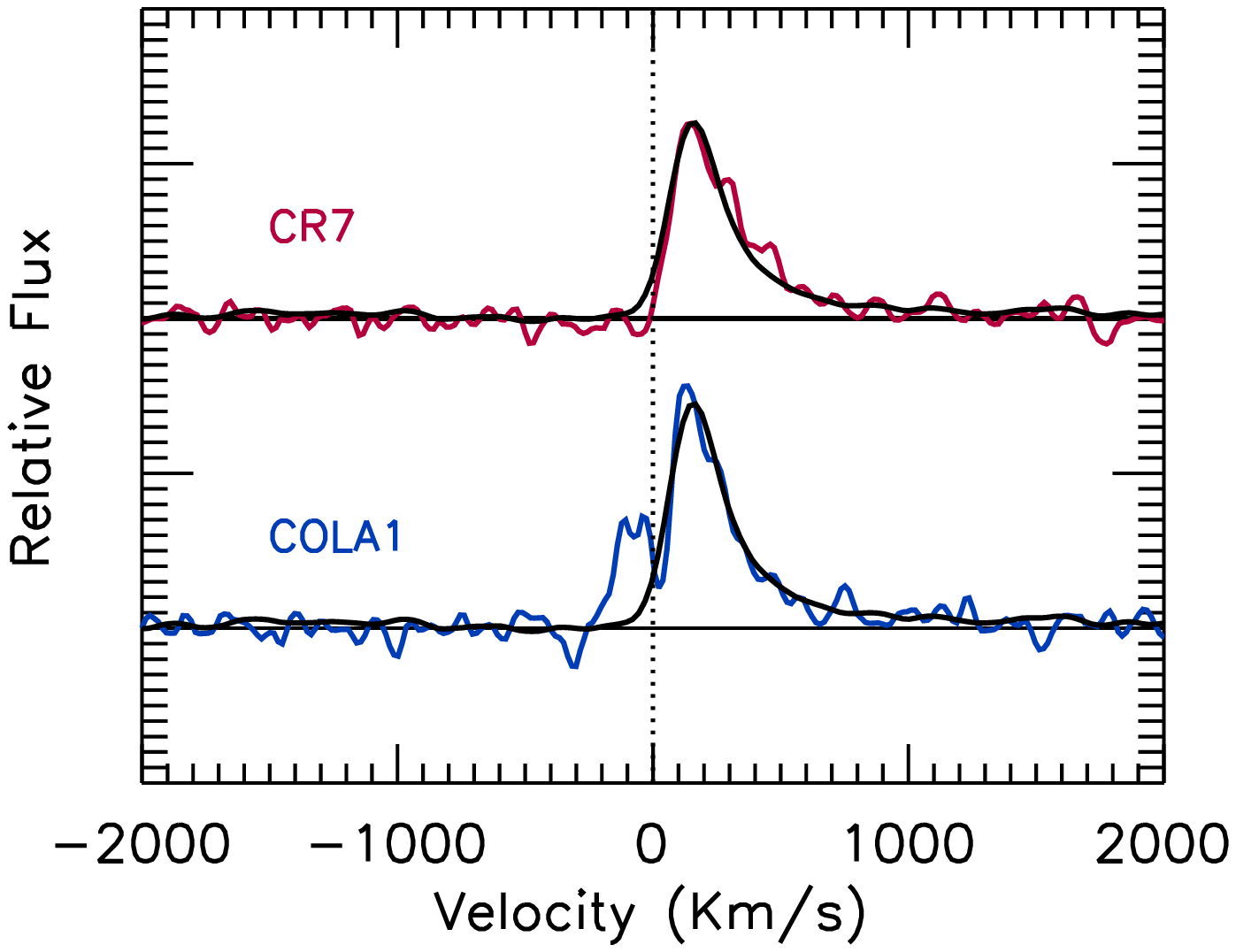}
\caption{Spectra of the two most ultraluminous LAEs.
{\it Bottom}\/: COLA1 {\it Top}\/: CR7. CR7's  velocity is in the
reference frame of the galaxy's HeII line, giving  
peak Ly$\alpha$  at $160~{\rm km\ s}^{-1}$. 
COLA1 is plotted with the peak  at the same offset.
Under this assumption the blue side extends to $\sim -200~{\rm km\ s}^{-1}$.
 Both objects also show a continuum
break across the line. The black overlay shows the composite
$z=6.6$ LAE profile of 
\citet{hu10}
normalized to match
the maximum in each of the objects.
\label{la_comp}
}
\end{figure}

We observed COLA1 with the Keck2/DEIMOS spectrograph on March
6th 2016, with the G830 grating
using a $1^{\prime\prime}$\  slit,  giving
a resolution measured from the sky lines
of 83 km s$^{-1}$ for the $z=6.6$ LAEs. Three
20-min exposures were obtained, dithered by $\pm 1.5^{\prime\prime}$ along
the slit, 
to obtain precise sky subtraction. The seeing was $0.6^{\prime\prime}$ and 
conditions were photometric. The data was reduced with our
standard pipeline 
\citep{cowie96}.
For detailed comparison with COLA1, we also observed CR7 in the same
configuration  but
we did not re-observe the fainter MASOSA. The 2-D 
spectra of COLA1 and CR7 are shown in Figure~2, compared
 with two of the most luminous LAEs from 
\citet{hu10}.
In Figure~\ref{la_comp}, we compare the line profile of COLA1 with
that of CR7 and also with the stacked spectra of lower-luminosity
LAEs at $z=6.6$\ from
\citet{hu10} 
with 
luminosities, $\log L({\rm Ly}\alpha) < 43.3~{\rm erg s}^{-1}$.

\begin{table}
\caption{Properties of COLA1 and CR7}
\begin{center}
\begin{tabular}{ccc}
\hline
& COLA1 & CR7 \\
 \hline
& & \\
Photometric& & \\

$g'$ & $27.26^{28.05}_{26.82}$ & $-27.42^{-28.38}_{-26.82}$ \\
\\
$i'$ & $27.10^{29.04}_{26.44}$ & $-27.38^{-30.17}_{-26.59}$ \\
\\
$z'$ & $25.39^{26.08}_{24.97}$ & $25.51^{26.33}_{25.04}$ \\
\\
NB$921$ & $23.50^{23.60}_{23.41}$ & $23.86^{24.01}_{23.73}$ \\
\\
$y'$ & $25.10^{25.53}_{24.79}$ & $24.71^{25.00}_{24.48}$  \\
& & \\
Spectroscopic& & \\
z(Ly$\alpha$)  & $6.593$  & $6.605$  \\  
$\log L({\rm Ly}\alpha)\ ({\rm erg\ s}^{-1})$ & $43.9$ & $43.8$\\
FWHM(Ly$\alpha$) (km\ s$^{-1}$) & $194$  & $247$ \\   
EW(Ly$\alpha$) (\AA) & $53$  & $99$  \\
\hline
\end{tabular}
\footnotetext{$3^{\prime\prime}$ diameter aperture magnitudes,
flux-calibrated by comparison with the COSMOS
magnitudes of 
\citet{capak07}
in the overlapping region. 
A minus sign denotes negative flux in the aperture and the number corresponds
to the absolute value of the flux. 
$1~\sigma$\ errors in the
band were determined by measuring the dispersion in blank field
positions with the same aperture, and are
NB921 26.1, $y'$ 26.3, $z'$ 26.2, $i'$ 27.3 and $g'$ 28.0, 
close to the predicted noise. The 
$1~\sigma$ range for each magnitude is shown with the upper and lower subscripts. }
\end{center}
\end{table}

The optical spectroscopic follow-up 
shows that COLA1 lies at a redshift of 6.593
based on the peak of the Ly$\alpha$ profile. 
However, more surprisingly,
it has a complex line profile with both blue and red components.
The flux of COLA1 is 1.4 times that of CR7 
based on the spectroscopy.
Even excluding the blue wing, COLA1 is still more luminous
than CR7, so the selection of COLA1 is not biased by its
unique profile. Indeed, the shape of the red wing is almost identical in the
two galaxies  and almost indistinguishable
from the stack of  lower-luminosity line profiles
given in \citet{hu10} (Figure~3).

%
%

%
\begin{figure*}
\centerline{\includegraphics[width=4.5cm,angle=0]{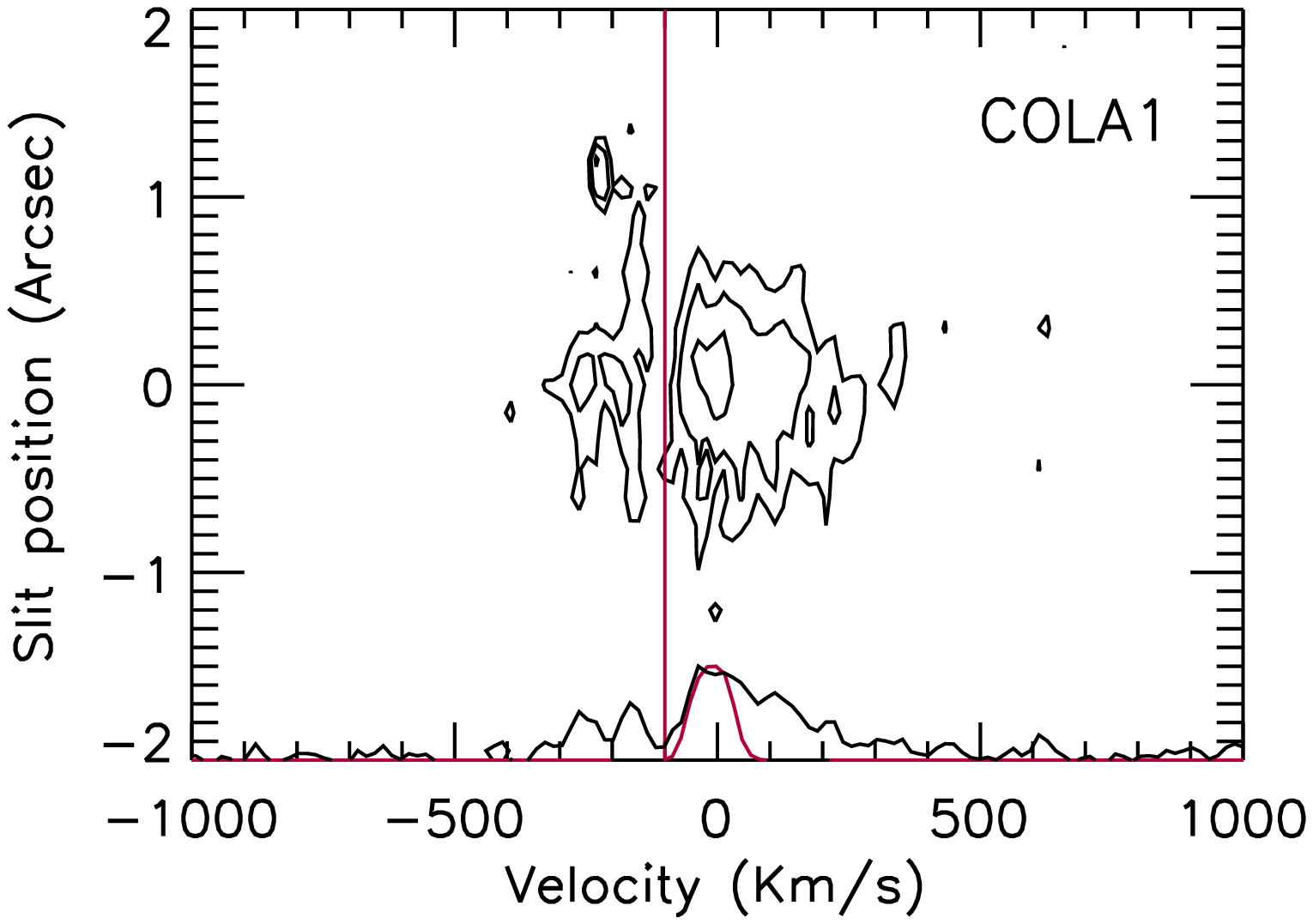}
\includegraphics[width=4.5cm,angle=0]{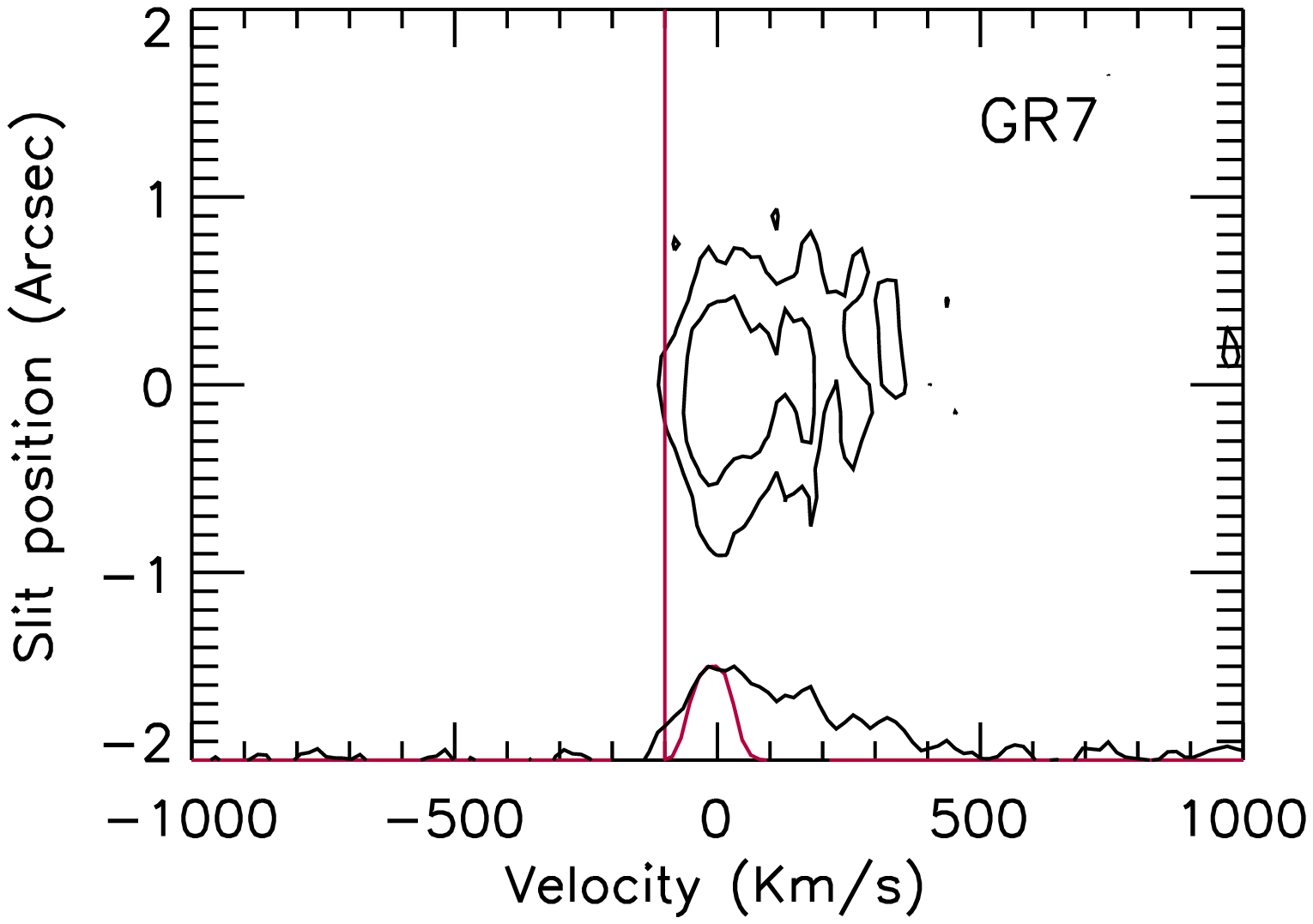}
\includegraphics[width=4.5cm,angle=0]{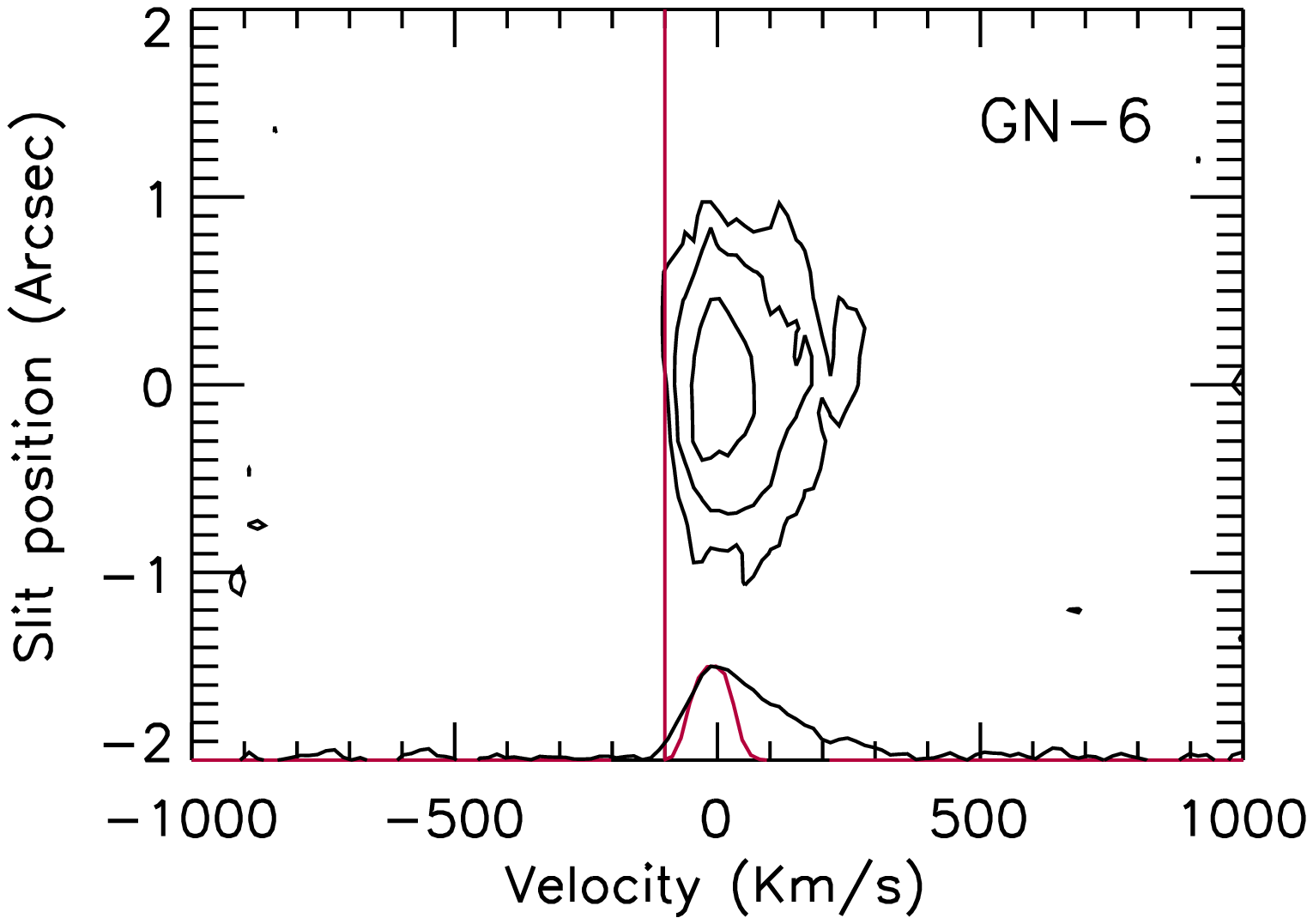}
\includegraphics[width=4.5cm,angle=0]{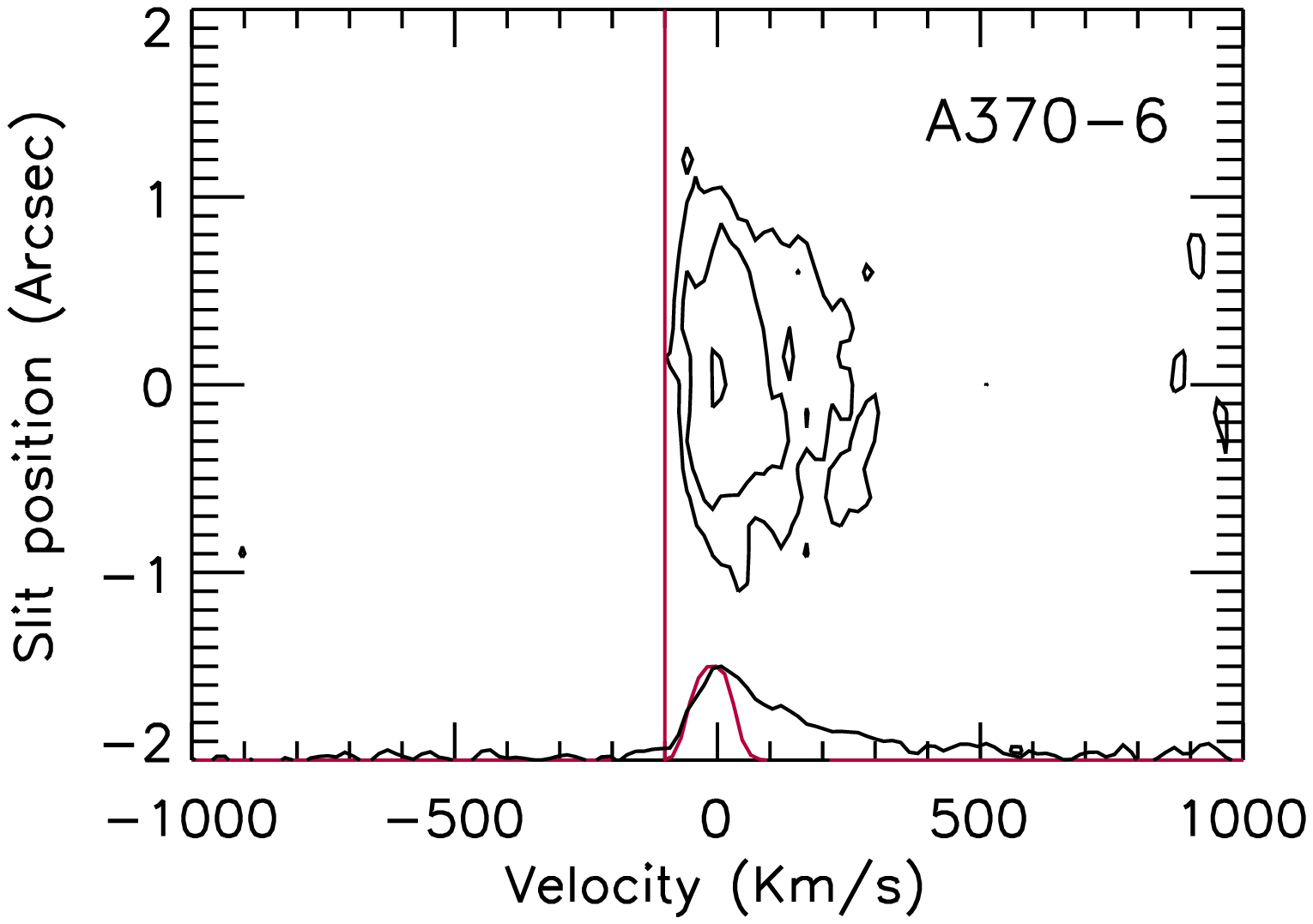}}
\caption{Two-dimensional spectra of the LAEs.
Ly$\alpha$ luminosity decreases from left to
right. COLA1 
($\log L({\rm Ly}\alpha) = 43.9\ {\rm erg\ s}^{-1}$) 
shows a clear blue extension, CR7
($\log L({\rm Ly}\alpha) = 43.8\ {\rm erg\ s}^{-1}$) 
does not show a clear sharp velocity edge, and the two lower
luminosity objects (GN-6 with 
$\log L({\rm Ly}\alpha) = 43.3\ {\rm erg\ s}^{-1}$) 
and A370-6 with 
$\log L({\rm Ly}\alpha) = 43.2\ {\rm erg\ s}^{-1}$) 
show the sharp razor-edge cutoff typical of  
lower-luminosity objects at this redshift.
In all cases the zero velocity is shown at the Ly$\alpha$
peak and the vertical red line marks $-100\ {\rm km\ s}^{-1}$. The
contours drop by a multiplicative factor of two. At the
bottom we show the line profile (black) and the spectral
resolution profile (red) centered at $0\ {\rm km\ s}^{-1}$.
\quad\quad\quad\quad\quad\quad\quad\quad\quad\quad\quad\quad\quad\quad\quad
\bigskip
\label{2d_countour}
}
\end{figure*}

The measured FWHM, corrected for the instrumental resolution,
is 194 km s$^{-1}$ in COLA1 compared to 247 km s$^{-1}$ in CR7.
(\citet{sobral15} 
give a Gaussian-fitted FWHM of 266$\pm$15 km s$^{-1}$ for CR7.)
Only the red side of COLA1 lies above the half maximum so the FWHM
relates to this component only. 
The measured rest-frame equivalent width of COLA1 is 53\AA\ and that of
CR7 99\AA, where in both cases we have measured the continuum
to  redward of the line. Both MASOSA and COLA1 are quite spatially
compact. In the narrow-band image
COLA1 has a measured FWHM of 0.93$^{\prime\prime}$ compared
with the locally measured PSF of $0.88\pm0.02^{ \prime\prime}$
giving an intrinsic FWHM of $0.3^{\prime\prime}$. In  $y\prime$
the local PSF is $0.77\pm0.01^{ \prime\prime}$ and
the COLA1 FWHM is 0.90$^{\prime\prime}$ giving an intrinsic FWHM of $0.46^{\prime\prime}$. This contrasts with CR7 which is quite diffuse \citep{sobral15}.
The present
data gives an intrinsic FWHM of $1.3^{\prime\prime}$ for CR7. There is no
sign of any \ion{N}{5} 1240~\rm\AA\ emission though this lies in a noisier
part of the spectrum where there are strong night sky lines.

At lower redshifts ($z=2.2$) \citet{konno15} 
suggest that almost
all Ly$\alpha$ emitters with $L({\rm Ly}\alpha) >10^{43.4}~
{\rm erg s}^{-1}$ are associated with AGN.
However, the narrowness of the Ly$\alpha$ line 
\citep[see][]{alexandroff13, matsuoka16}
the spatial extension
of the galaxy and the absence of \ion{N}{5} 1240\AA\ (though the latter
two constraints are weak) combine to suggest that COLA1 
is primarily powered by star formation rather than AGN activity. However,
both COLA1 and CR7 could have AGN contributions to their emission.

\vskip 1in
\section{Physical interpretation}
\label{secphys}

It seems, therefore, that the most luminous LAE ever detected in
the epoch of reionization has a unique and unexpected line profile.
The simplest explanation for this surprising result is that COLA1 lies in a
highly ionized region of gas, increasing the Ly$\alpha$ forest transmission
and thereby allowing us to see the blue side of the Ly$\alpha$ line profile.
Alternatively, the galaxy could be moving at a velocity of several hundred km
s$^{-1}$ with respect to the IGM, so that the Ly$\alpha$ profile has been
moved redward of the effects of IGM scattering. The compactness of the galaxy
and the spatial alignment of the blue and red wings in the 2-D spectrum makes
it unlikely the velocity structure is caused by a galaxy merger.

\citet{matthee15} have argued that there is little evolution in the luminosity
function of the most luminous LAEs at these redshifts, suggesting that these
objects lie in large HII regions and protect themselves from any changes in
IGM neutrality.  This would be consistent with complex Ly$\alpha$ profiles
being seen only in the most ultraluminous LAEs.

At low redshifts ($z=2-3$) about 30$\%$ of LAEs
show multi-component profiles \citep{kulas12}.
Kulas et al.\ argue that these galaxies are not strongly
affected by IGM opacity so that this fraction could also
be applicable to the high-redshift galaxies enclosed in
giant HII regions (though of course the properties of
the high-redshift galaxies may be very different). Assuming 
that COLA1, CR7 and MASOSA lie in this class
then we do indeed see one galaxy in three having a complex
profile, which would be crudely consistent.
If this is the correct interpretation, then more complex profiles may 
be present in a fraction of the most ultraluminous LAEs.

COLA1's velocity structure appears very similar to that of 
multicomponent LAEs at lower redshift though with
lower velocities \citep[e.g.][]{kulas12, quider10}.
The peak in the blue side has a maximum that is 
0.44 times that of the red side and the velocity
separation between the red and blue peaks 
is 215 km s$^{-1}$. Our assumed velocity
offset places the trough between the peaks at 25 km s$^{-1}$,
close to the near-zero velocity seen in most 
low-redshift LAEs in \citet{kulas12}.

As discussed in \S 1, 
interpreting the LAE profiles is complicated \citep[e.g.][]{verhamme12}. 
 The velocity separation of the peaks and the redside
velocity can be produced with an expanding shell with
a velocity of $\sim 50~{\rm km\ s}^{-1}$ but such a model
would predict a much stronger red-to-blue ratio than is
seen in COLA1, suggesting we also need significant infall.
We postpone  more detailed discussion until an
an exact determination of the absolute velocity scale is obtained.
However, if COLA1 does have a strong inflow this
 might suggest that the galaxy is in an early formation phase and
has a low metallicity.  This would be consistent with  the \citet{sobral15}
interpretation of  CR7 
as having a young, metal-poor stellar population.

If we accept that the complex line's visibility is a result of HII region ionization, then 
we may be able to draw useful inferences about the IGM and  the escape 
of  ionizing photons from the galaxy. 
Unfortunately the reduction in the Ly$\alpha$ flux and the scattering of the blue
side by the presence of the HII region depend on
a very large number of parameters.  The HII region size 
depends on the total number of ionizing photons from
the galaxy and inversely on $x_{\rm IGM}$, the IGM neutral hydrogen fraction.
\citet{haiman02} gives
the size as

\begin{equation}
R_S = 0.45 \,x_{\rm IGM}^{-1/3} \,\left (N_{ion} \over {5\times10^{69}} \right )^{1/3} \,\,\rm{Mpc}\ .
\end{equation}\

\noindent
Here R$_s$ is in proper Mpc and $N_{\rm ion}$ is the total number of ionizing
photons released by the galaxy. $N_{\rm ion}$ depends on the star-formation
rate of the galaxy, its initial mass functions, the ionizing
photon escape fraction ($f_{\rm esc}$), and the age of the galaxy. It is correspondingly an extremely
uncertain quantity. Nevertheless, the HII region radius is generally large enough
that we can largely ignore the effects on the Ly$\alpha$ line of the radiation 
damping wings of the
IGM outside the HII region unless the
IGM is substantially neutral \citep{miralda98, haiman02}.

In this case the amount of scattering in  the
Ly$\alpha$ profile is dependent only on the residual density of neutral gas
in the HII region \citep{haiman02, haiman05}.
This 
affects only the blue side of the profile where the gas lies at lower
redshifts than the galaxy. The fractional neutral hydrogen
in the HII region is given by \citet{cen00} 
as 

\begin{equation}
x_{\rm HII} = 8\times10^{-4} C_{\rm HII}\,\left (r \over {\rm Mpc}\right )^{2} \,\left (\dot{N}_{ion}\over 10^{54}~{\rm s}^{-1}\right )^{-1}\ . 
\end{equation}\

\noindent
Here $C_{\rm HII}$ is the clumping factor, $r$ is the radial distance from the galaxy,
and $\dot{N}$$_{ion}$ is the rate at which the galaxy releases  ionizing photons into
the IGM. Translated simply, this predicts a large optical depth even for relatively
low values of $\dot{N}$$_{ion}$. 
\citet{haiman02} has values of $\tau_{\rm HII}$ near
100 for $\dot{N}$$_{ion} \sim 10^{54}$ s$^{-1}$, characteristic of a
moderate luminosity LAE with a high escape fraction. However, the effects of structure must be taken
into account since the transmission is dominated by lower-density regions
and this significantly reduces the effective optical depth. With plausible
assumptions, \citet{haiman02} 
showed that this effect reduces the optical depth by
almost three orders of magnitude, placing an optical depth of one on the
immediate blue side of the LAE for a galaxy with this value of $\dot{N}$$_{ion}$.
Thus it is plausible that a change in the LAE properties occurs when 
$\dot{N}$$_{ion}$ rises significantly above $10^{54}$ s$^{-1}$ and the
HII region becomes effectively thin.
If we assume this change occurs at around the Ly$\alpha$ luminosity of MASOSA
($\log L({\rm Ly}\alpha)=43.4$; Sobral et al.\ 2015) 
we can convert this to an
ionizing photon release of 
$\sim 1.6\times 10^{46}f_{\rm esc}$,
assuming
the case B conversion to $L({\rm H}\alpha)$ and a Salpeter IMF 
\citep[e.g.][]{matthee15}.
The two match for 
$f_{\rm esc} \sim 0.01$ 
but the significant
uncertainties in the computation of the optical depth arising
from the HII region must be borne in mind.
 
If this is correct  we would expect to see a change in the structure of the
lines at the transition point. Below the critical $\dot{N}$$_{ion}$ the effects
of  HII absorption should produce a sharp razor-edge cutoff at all positions
on the galaxy, as indeed is seen in the lower-luminosity LAEs (Figure 4) where
the cutoff essentially corresponds to the instrument profile. At higher
luminosities the cutoff is just due to the profile emerging from the galaxy
and can show variation across the galaxy, as is seen in both COLA1 and CR7.

\section{Summary}
\label{secsum}

The discovery of COLA1 at $z=6.593$ --- the most luminous LAE ever detected 
near the reionization epoch and the highest redshift LAE 
to show a multi-component structure with a blue wing -- 
has the potential to significantly improve our understanding of 
the ionization of the IGM by galaxies at $z=6.6$. We suggest
that the most luminous LAEs lie in HII regions which are sufficiently
highly ionized by the galaxy to allow unscattered transmission of
the blue side of the Ly$\alpha$ line. The transition Ly$\alpha$ luminosity at which this occurs 
allows an estimate of the escape fraction of  ionizing photons
from the galaxy. If this interpretation is correct then a substantial
number of the most ultraluminous LAEs may show complex line profiles,
and modeling of the profiles should allow us to determine the kinematic
structure of the galaxies. Large samples of ultraluminous LAEs should
soon be available from wide-field surveys with HSC on Subaru and should
allow us to study these issues in detail.

\acknowledgements

We gratefully acknowledge support from NSF grants
AST-1313309 (L.~L.~C.) and AST-1313150 (A.~J.~B) and
the University of Wisconsin Research Committee with funds
granted by the Wisconsin Alumni Research Foundation (A.~J.~B.).
We would like to thank John Silverman and Guenther Hasinger
for providing the reductions of the HSC data on the COSMOS field
and Zoltan Haiman and the anonymous referee for their reviews
of the first draft of the manuscript.
We acknowledge the cultural significance that the summit of 
Mauna Kea has for the indigenous Hawaiian community.

\end{document}